# Complex dynamics and configurational entropy of spatial optical solitons in nonlocal media

Claudio Conti[1]

[1] Research Center "Enrico Fermi", Via Panisperna 89/A, Rome (Italy)

and Research Center Soft INFM-CNR, University "La Sapienza," Piazzale Aldo Moro 2, 00185, Rome (Italy)

Marco Peccianti[2] and Gaetano Assanto[2]

[2] N*oo*EL - Nonlinear Optics and OptoElectronics Laboratory

Department of Electronic Engineering and INFM-CNISM, University "Roma Tre"

Via della Vasca Navale, 84 - 00146, Rome (Italy)

Intense light propagating in a nonlinear medium can generate an ensemble of interacting filaments of light, or spatial solitons. Using nematic liquid crystals, we demonstrate that they undergo a collective behavior typical of complex systems, including the formation of clusters and sound-like vibrations, as well as the reduction of the configurational entropy, controlled by the degree of nonlocality of the medium.

Complexity deals with those phenomena which, emerging from the interaction of several objects, despite their individual features, share a universal character and are observable in very different systems, including for example spin glasses,[1] life science,[2] laser physics and pattern formation.[3] The collective behavior depends on global parameters such as the total number of objects and the mutual interaction, rather than the constitutive entities. Some recent investigations indicate that intense light propagating in a nonlinear medium can be assimilated to an ensemble of interacting particles, i. e. filaments of light or spatial solitons.[4-7] Up to thousand filaments have been observed,[8] featuring limited distortion in a cloud,[9] self-healing after an obstacle[10] and super-continuum generation.[11]

In this Letter we investigate various features of propagating light filaments which typically characterize complex systems, e. g. clustering, dynamic phase transitions and quenched vibrations. Indeed, an ensemble of solitons -mutually interacting as they propagate- can be assimilated to molecules evolving in time in a colloidal suspension or "soft-matter," as for instance polymeric solutions, liposomes, foams, etc. (see e.g. Ref. 12 and references therein). At a microscopic level, soft-matter can be described in terms of particles which interact with a given potential energy, the latter depending on both molecular composition and solution properties. In such "complex" materials, phase changes are related to the formation of aggregates or clusters as characteristic parameters -such as density or interaction range- vary throughout the system. These phase transitions are not purely thermodynamic but rather "dynamic."[13-16] Hereby we study dynamic phase transitions (DPT) of a light beam occurring as it propagates in nematic liquid crystals (NLC)- and featuring the formation of filament clusters. We will show that this phenomenon is accompanied by changes in both the noise figure and the vibrational spectra (which can be quantified by appropriately defining a *configurational entropy*), assimilating light to a complex medium.

In our case, a DPT concerns the role of index fluctuations in the distribution of an ensemble of interacting solitons (see Ref. 17), it denotes the transition between two regimes: one in which number and distribution of soliton clusters is strongly affected by director fluctuactions and one where the clusters are "quenched."

In the experiments we employed a planar cell filled with the nematic liquid crystal E7, as detailed in Ref. 5 and temperature stabilized at 26°C. As the beam propagates forward, light scattered above the cell enables us to monitor the optical dynamics along longitudinal and transverse coordinates $z$ and $x$, respectively, by the use of a CCD camera and a microscope. The optically induced index perturbation results from molecular reorientation, as the electric field **E** tends to reorient the NLC axial director $\hat{\mathbf{n}}$ towards the linear polarization of light.[18,19]

When an intense extraordinary-polarized (i.e. with **E** in the plane **k**-$\hat{\mathbf{n}}$) elliptic beam (its width along $x$ being much larger than its "depth" along $y$) excites the sample, light propagates along $z$ generating filaments through modulational instability (MI), as apparent in Figure 1 for various voltages applied across the cell thickness (75μm along $y$).[5]

Due to the finite elasticity of NLC, each filament is associated to an index perturbation Δn spatially extending far beyond the filament cross-section, i. e. to a non local response. Furthermore, the applied bias governs not only the director distribution $\hat{n}(x,y)$ in the absence of light, but also the degree of filament-to-filament interaction by contrasting or promoting molecular reorientation.[20] Hence we are able to control the degree of non-locality and tune the system through dynamic phases, each of them identified by a peculiar regime of filament propagation.[17] For a small non locality, corresponding to voltages around and above 2.5V, solitons are well separated in the transverse coordinate, exhibit a negligible mutual interaction and propagate almost independently. This is apparent in the last two panels of figure 1 (3.0 and 2.5 V). Interference between filament pairs is averaged out due to their large number and propagation in a randomly fluctuating liquid. Conversely, for a strong interaction when a lower bias enhances non locality, [20] filaments form multiple and disordered clusters, as in figure 1 for bias V<2.5V.

Since director fluctuations introduce randomness in the cluster distribution, we are addressing the interplay between noise-induced disorder and nonlocality. Specifically, the observations above are properly interpreted in the frame of complex systems, where the potential is replaced by a "potential energy landscape" (PEL), i.e. a multidimensional potential energy surface depending on each particle position.[21] As a PEL can exhibit many minima (the larger their number the more complex the dynamics), a DPT can be defined as a PEL deformation which makes the system switch from simple (isolated minima) to complex dynamics (several minima) and vice-versa. Numerical simulations confirm the scenario. [17]

Since a PEL is hard to visualize, from the experimental data we can obtain a simple portrait of it by extracting, digitalizing and recording the position of each filament across the transverse coordinate $x$. A string of binary bits –its length determined by the number of CCD pixels along $x$- can be associated to the transverse distribution of the optical wave-front at a given propagation distance $z=Z$ from the input (we take Z=3.0mm, i.e. tens diffraction lengths for each filament), each bit "1" ("0") corresponding to the presence (absence) of -at least- one soliton. Iterating the procedure for several noise realizations, i.e. distinct snapshots at different times (separated by 100ms), a PEL representation corresponds to the probability of a bit "1" (i.e. a filament) in a given position. [2] Figure 3(A-C) graphs this distribution for three voltages. For biases ≥1V (see Fig. 3A-B), the PEL is highly disordered and the number of peaks varies with voltage; conversely, at low bias (see Fig. 1 and Fig. 3C) the peaks correspond to two dominant clusters.. In our NLC system, at high bias V several energetically-equivalent attractors (i.e. one for every soliton cluster configuration with comparable probability) are present, whereas at low V a dominant minimum corresponds to two clusters. Furthermore, as clustering must correspond to a modified degree of disorder, the latter can be quantified by the Shannon entropy of the strings associated to filaments. After identifying the soliton positions as the intensity maxima (i.e. calculating 1st and 2nd derivatives with respect to $x$ and filtering out the noise by retaining only points with intensities >50% of the maximum), those location can be averaged on all snapshots in order to determine the probability $p_i$ of a filament to be at the i-*th* pixel (i = 1,… ,1024), yielding the corresponding configurational entropy (see e.g. Ref. 2) as $\sum_{i=1}^{1024}[-p_i \log(p_i)]$.

Fig. 3D displays such entropy at two distances Z, clearly demonstrating a strongly reduced disorder at low voltages.

On the one hand, when the complex system evolves through different (equivalent) energy minima it is subject to pronounced oscillations (driven by noise in the medium); on the other hand, in the low-voltage phase only one minimum is present and fluctuations are quenched. At high V each soliton undergoes an independent Brownian motion and -in essence- the fluctuations measure the noise in the medium. In the disordered phase (intermediate V) the fluctuations are more pronounced because of the several local PEL minima routinely "visited" by the filaments. At the DPT (low V), two dominant clusters are created and the dynamics "slows down" with noise reduction: the system settles in a deep PEL minimum (i.e. in a dominant attractor).

It is worthwhile stressing that, at variance with the field entropy considered in Ref. 22 and increasing as solitary wave generation occurs triggered by modulational instability, the configurational entropy gives an indication on the number of cluster states populated by the soliton ensemble in the presence of disorder.

Finally, we also report on the evolution of the intensity spectra with the degree of interaction. These spectra can be obtained by Fourier transforming and time-averaging the acquired images. When all filaments propagate independently (e.g., Fig. 1 at high V), the transformed pattern essentially consists of a series of peaks with negligible width along $k_z$, the image being substantially *z*-independent (see Fig. 4(A)). When the "particles" interact (lower bias), their spectra are characterized by oblique lines. This is visible in figures 4(B-D) and resembles the vibrational spectra in glassy systems (the z-coordinate here corresponding to time there). The intensity profile can be taken as a coarsely-grained density of filaments/particles and the oblique lines indicate sound-like vibrations in the *x-z* plane, corresponding to an increased interaction range as predicted by the general theory of random matrices.[17,23]

In conclusion, we investigated the interplay between non locality and disorder in the propagation of optical spatial solitons in nematic liquid crystals. Dynamic phase transitions of light in such a complex system were demonstrated for the first time.

C.C. thanks L. Angelani, L. Lezzi and G. Ruocco for enlightening discussions.

**Figure captions**

Fig. 1. (Color online) Filament evolution in the plane *x-z* at various biases V, for a 260mW input beam (1064nm) of waist 500 μm along *x* and 5 μm along *y* injected into an NLC cell kept at 26°C. The filaments appear slanted because of walk off, which also depends on applied voltage.

Fig. 2. (Color online) Intensity distribution versus bias: time averaged intensity distribution (over 50 snapshots separated by 100ms) at Z=3mm for various biases V (see also Fig. 18 in Ref. 17).

Fig. 3. (Color online) (**A-C):** Landscape (PEL) sections as the probability of finding a filament at a given position, obtained from 50 camera snapshot for three voltages. (**D**): Shannon entropy of the strings vs voltage, determined in Z=2 and Z=3mm, respectively.

Fig. 4. (Color online) Vibrational spectra: time-averaged spatial spectra of the intensity for various biases, with the formation of sound-like spectra (oblique lines in **C-D**) following the DPT.

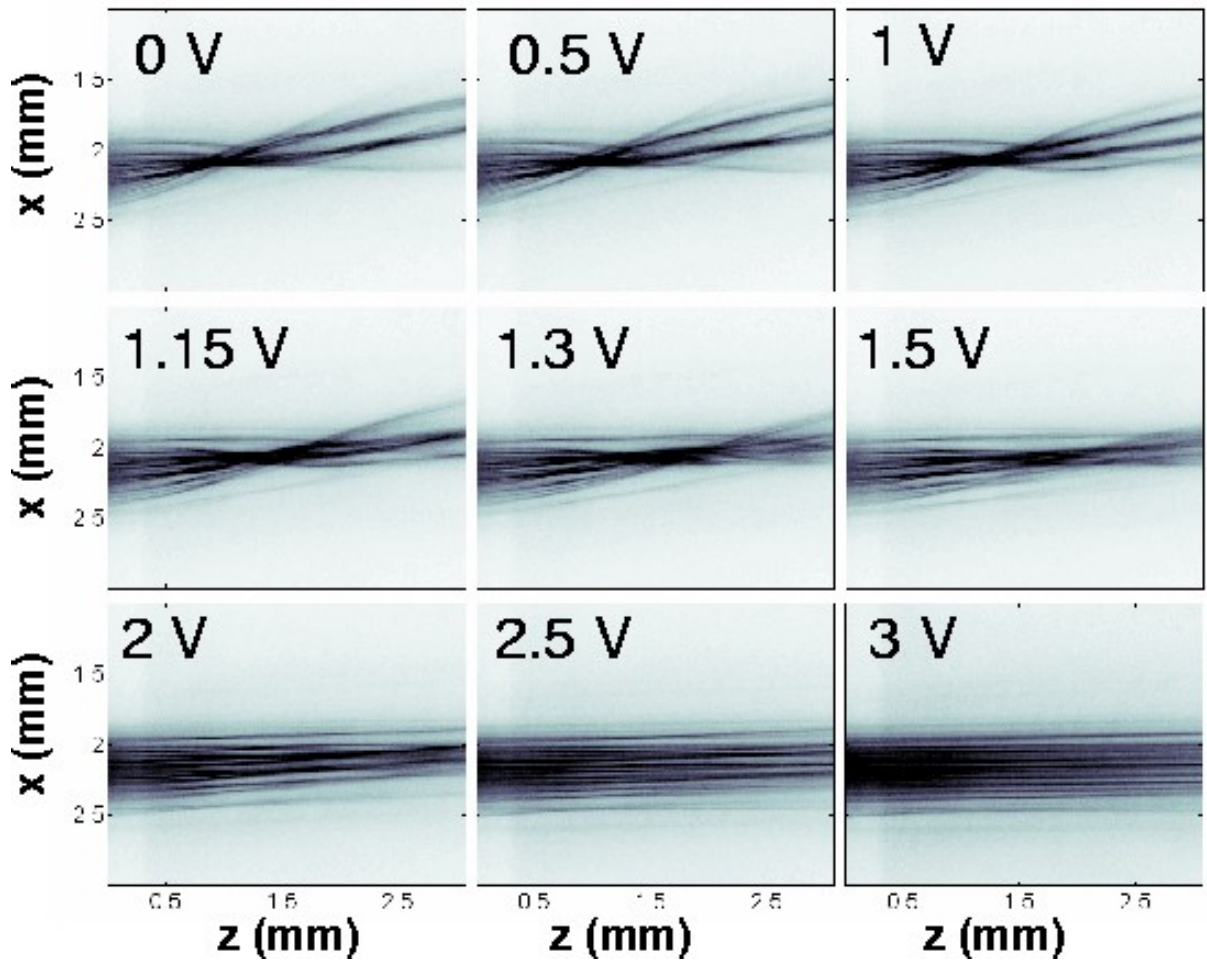

FIGURE 1

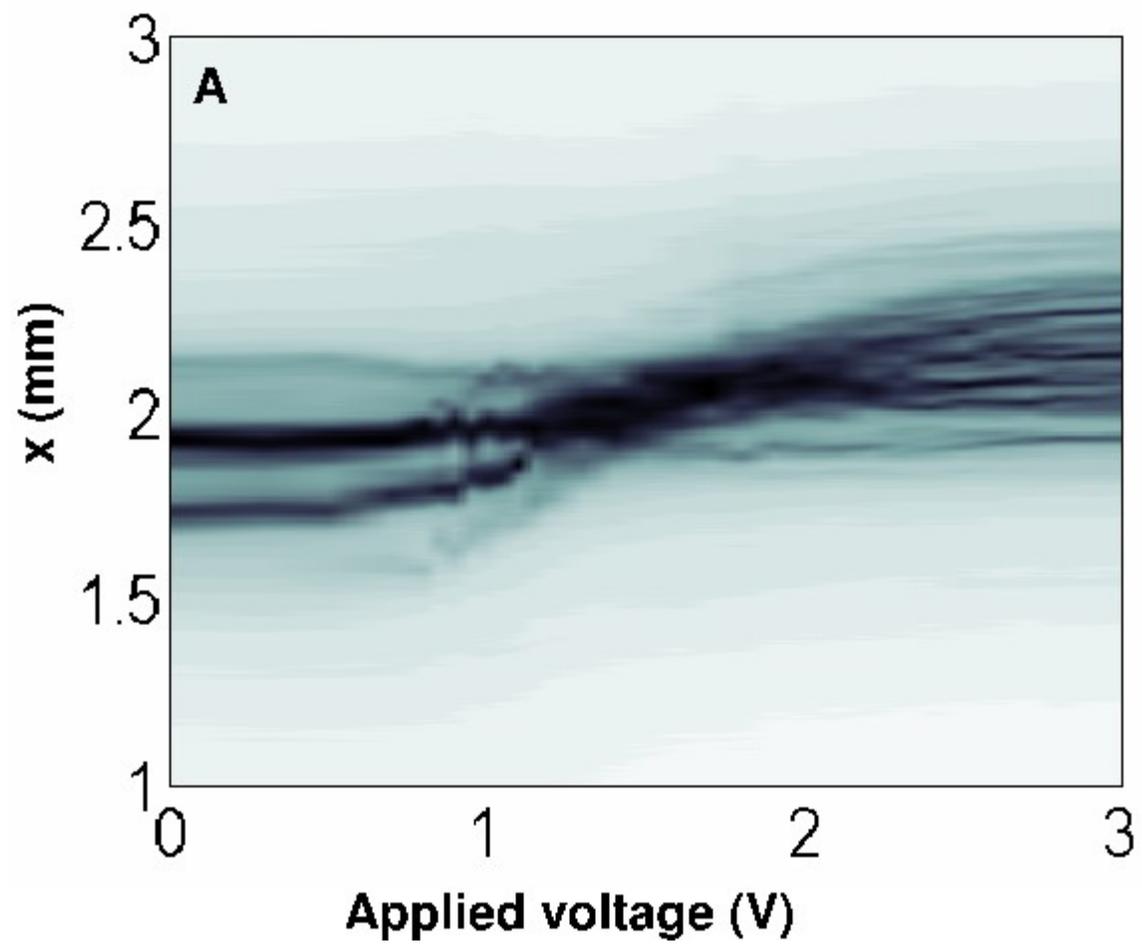

FIGURE 2

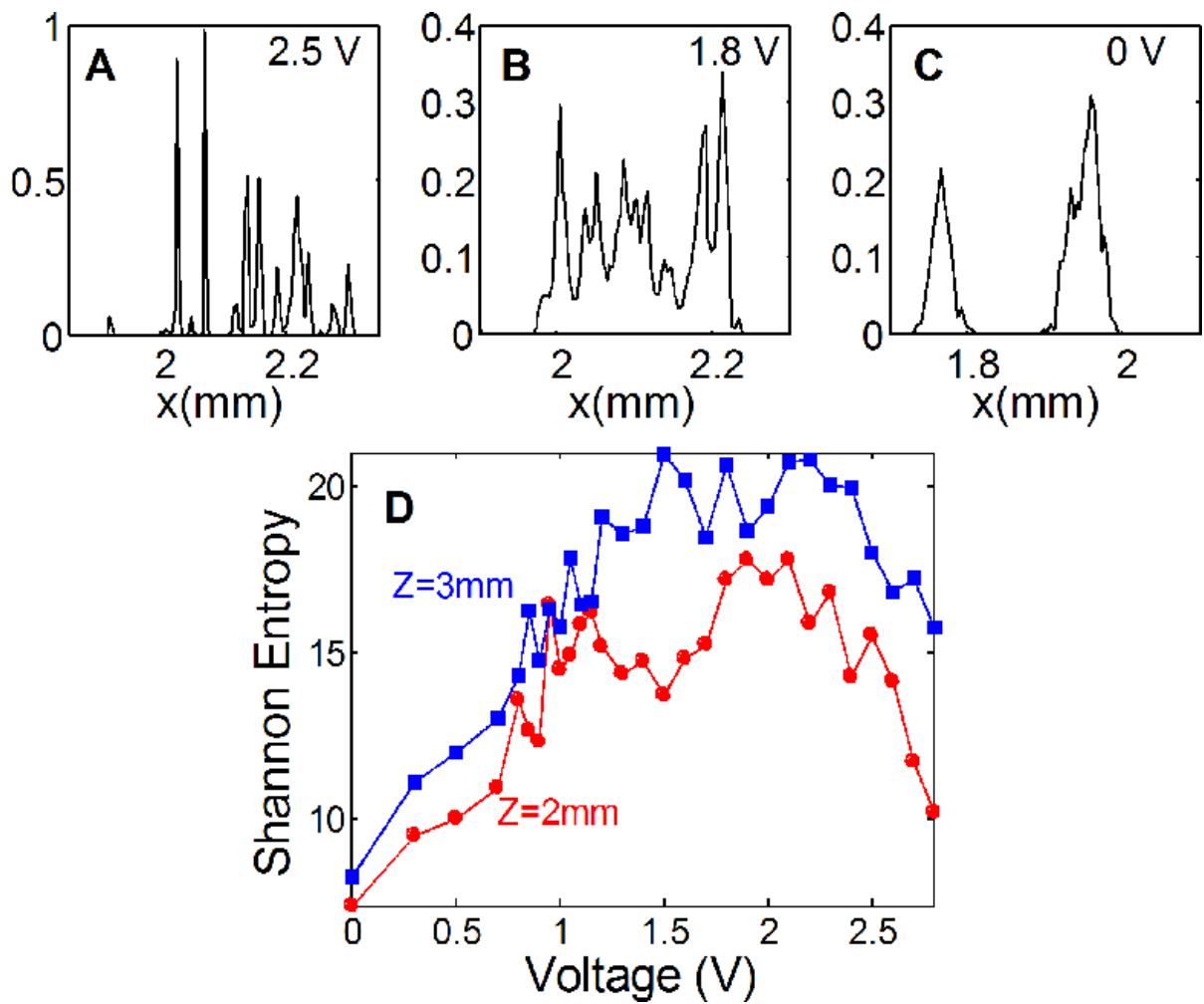

FIGURE 3

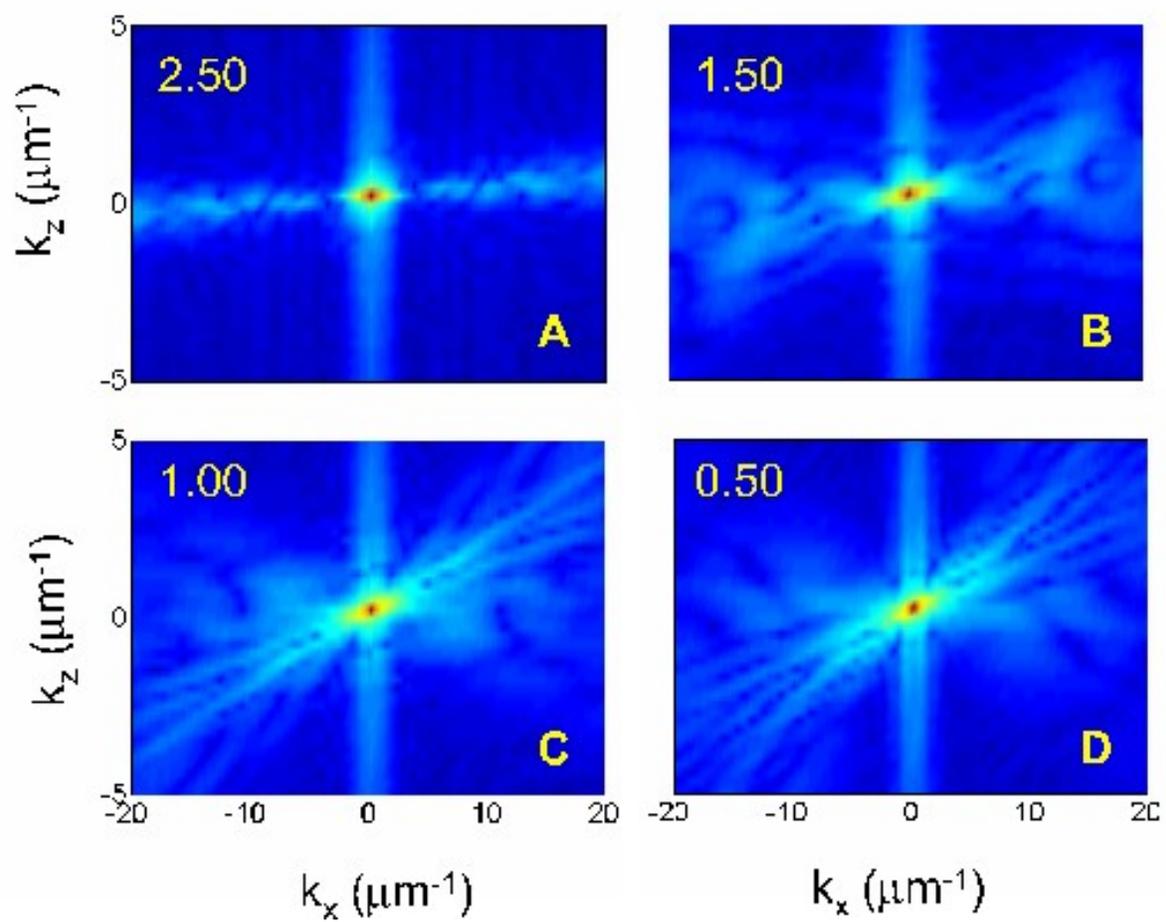

FIGURE 4